\documentclass[11pt]{article}
\usepackage[utf8]{inputenc}
\usepackage{amsfonts,epsfig}
\usepackage[hyphens]{url}
\RequirePackage{color}
\usepackage{hyperref}
\hypersetup{
	colorlinks=true,
	linkcolor=black,
	citecolor=black,
	filecolor=black,
	urlcolor=black,
}
\usepackage{breakurl}
\usepackage{comment}
\graphicspath{{fig/}}
\usepackage{relsize}
\usepackage{fancyvrb}
\usepackage{amssymb}
\usepackage{geometry}
\usepackage{sectsty}
\usepackage{caption}
\usepackage{subcaption}
\usepackage{amsmath}
\usepackage{amsthm}
\usepackage{bbm}
\usepackage{extarrows}
\usepackage{fancyvrb}
  \usepackage{tikz}
 \usetikzlibrary{shapes,arrows,chains}
 \usetikzlibrary[calc]
 \usetikzlibrary{bayesnet}
\usepackage[defaultsans, scale=0.93]{opensans} 
\usepackage[T1]{fontenc}
\usepackage{inconsolata}

\DeclareMathOperator{\R}{\mathbb{R}}
\DeclareMathOperator{\N}{\mbox{N}}
\usepackage[normalem]{ulem}
\sectionfont{\sffamily\bfseries\upshape\large}
\subsectionfont{\sffamily\bfseries\upshape\normalsize} 
\subsubsectionfont{\sffamily\mdseries\upshape\normalsize}
\makeatletter
\renewcommand\@seccntformat[1]{\csname the#1\endcsname.\quad}
\makeatother
 \usepackage[round]{natbib}
 \usepackage{mathtools}

\makeatletter
\def\@maketitle{%
	\begin{center}%
		\let \footnote \thanks
		{\large \@title \par}%
		{\normalsize
			\begin{tabular}[t]{c}%
				\@author
			\end{tabular}\par}%
		{\small \@date}%
	\end{center}%
}
\makeatother

\title{
	\bf Bayesian Aggregation   \vspace{.1in}
}
\author{
	Yuling Yao\footnote{Department of Statistics, Columbia University.  \texttt{yy2619@columbia.edu}}}
\date{  }

\begin{document}\sloppy
	\maketitle
\thispagestyle{empty}

\begin{abstract}
A general challenge in statistics is prediction in the presence of multiple candidate models or learning algorithms.  Model  aggregation tries to combine all predictive distributions from  individual models, which is more stable and flexible than single model selection.
 In this article we describe when and how to aggregate models under the lens of Bayesian  decision theory. 		
Among two widely used methods, Bayesian model averaging (BMA) and Bayesian stacking, we compare their predictive performance,  and 
review their  theoretical optimality,   probabilistic interpretation, practical implementation, and extensions in complex models.  

\end{abstract}

\section{From Model Selection to Model Combination}
  Bayesian inference provides a coherent workflow for data analysis, parameter estimation, outcome prediction, and uncertainty quantification. However, the model uncertainty is not automatically calibrated: the posterior distribution is always conditioning on the model we use,   in which the true data generating mechanism is almost never included.    No matter if viewed from the perspective of a group of modelers holding different subjective beliefs, or a single modeler revising belief models through the routine of model check and criticism, or the need of expanding plausible models for flexibility and expressiveness,  it is common in practice to obtain a  range of possible belief models. 

In Section \ref{sec_decision}, we review Bayesian decision theory, through which the model comparison, model selection, and model combination are viewed in a unified framework. The estimation of the expected utility depends crucially on how the true data generating process is modeled, and is described by different $\mathcal{M}$-views in Section \ref{sec_M_views}.   We compare Bayesian model averaging and leave-one-out (LOO) based Bayesian stacking in Section \ref{sec_compare}, which corresponds to the $\mathcal{M}$-closed and  $\mathcal{M}$-open view respectively. To explain why these methods work, we discuss related asymptotic theories in Section \ref{sec_theory}. In Section \ref{sec_practice}, we investigate the computation efficiency, and  demonstrate an importance-sampling based implementation in \texttt{Stan} and  \texttt{R} package \texttt{loo}.  We also consider several generalizations in non-iid data.

  \begin{figure}[t]
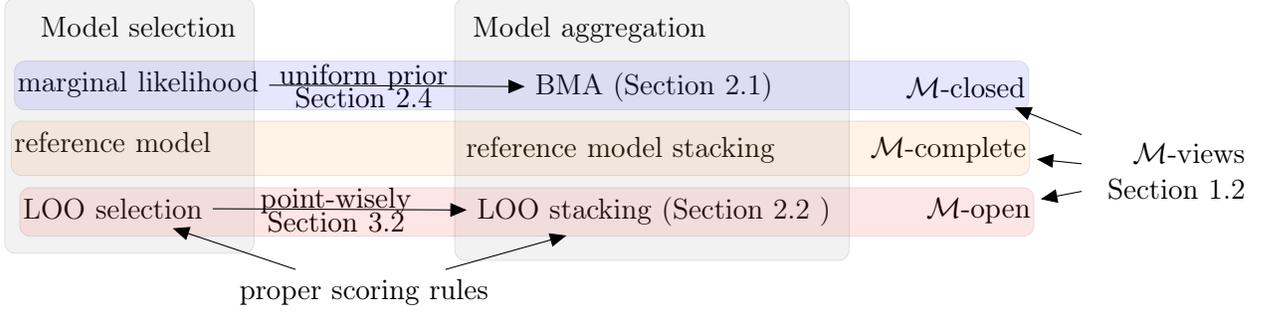

	\centering
	\tikz{
		\node (p_s)  {Bayesian Decision Theory Framework For Model Assessment};%
		\node[  below =of p_s ,xshift= -3cm, yshift= 0.9cm] (m_s)  {Model selection}; %
		\node[  below =of p_s ,xshift=  3cm, yshift= 0.9cm] (m_a)  {Model aggregation}; %
		\node[  below =of m_a ,xshift=  7.8cm, yshift=-0cm] (M_v)  {
			 \begin{tabular}{r}
			$\mathcal{M}$-views \\
			 Section \ref{sec_M_views}\\
			\end{tabular}}; %
			\node[  below =of m_a ,xshift=  4.77cm, yshift=-0cm](M_com){$\mathcal{M}$-complete}; %
			\node[   below =of m_a ,xshift=  5cm, yshift=0.8cm] (M_col)  {$\mathcal{M}$-closed}; %
			 \node[  below =of M_com ,xshift=  0.4cm, yshift=0.8cm] (M_o)  {$\mathcal{M}$-open}; 
			 
		\node[  left =of M_col ,xshift=  -0.5cm] (BMA)  {BMA (Section \ref{sec_BMA}) }; %
	    \node[  left =of M_com ,xshift=  0cm] (ref)  {reference model stacking}; %
		\node[  left =of M_o ,xshift=  0cm] (stack)  {LOO stacking  (Section \ref{sec_stacking} )}; %
		
				\node[  below =of m_s ,yshift=  0.8cm] (ml)  {marginal likelihood}; %
		\node[   below =of m_s ,yshift=  0cm , xshift= - 0.34 cm] (ref_s)  {reference model}; %
		\node[   below =of ref_s ,yshift=  0.65cm, xshift= - 0 cm] (loo)  {LOO selection}; %
     \node[  right  =of ml ,xshift=  -1cm, yshift=0.1cm]  (ss1)  {uniform prior};
          \node[  below  =of ss1 , yshift=1.3cm]  (ss2)  {Section \ref{sec_connect}}; 
             \node[  right  =of loo ,xshift=  -0.5cm, yshift=0.1cm]  (ss3)  {point-wisely};
          \node[  below  =of ss3 , yshift=1.3cm]  (ss4)  {Section \ref{sec_novel}};   
                    \node[  below  =of p_s , yshift=-2.6cm]  (pps)  {proper scoring rules};   
				\plate [inner sep=-0.1cm,yshift=0.1cm, fill=blue, opacity=0.1] {plate1}{(M_col)(ml)}{}; 
            	\plate [inner sep=-0.1cm,yshift=0.1cm,fill=red, opacity=0.1] {plate2}{(M_o)(loo)}{}; 
            	 \plate [inner sep=-0.1cm,yshift=0.1 cm, fill=orange, opacity=0.1] {plate3}{(M_com)(ref_s)} {} ; 
            \plate [inner sep=.1cm, fill=gray,  opacity=0.1] {plate4}{(m_s)(loo)} {} ; 
            	   \plate [inner sep=.1cm,  fill=gray,  opacity=0.1] {plate5}{(m_a)(stack)} {} ; 
            	 \edge  {ml} {BMA} 
            	  \edge  {loo} {stack} 
            	              	  \edge  {M_v} {M_o} 
            	       \edge  {M_v} {M_com} 
            		  \edge  {M_v} {M_col} 
                   \edge  {pps} {stack} 
                    \edge  {pps} {loo} 
	}\caption{\em The organization and connections of  concepts in this paper.}\label{fig_graph}
\end{figure}

 \subsection{The Bayesian Decision Framework for Model Assessment}\label{sec_decision}
  We denote  $D=\{ (y_1, x_1), \dots ,  (y_n, x_n) \}$   a sequence of observed outcomes $y\in \mathcal{Y}$ and  covariates $x \in \mathcal{X}$.
 The unobserved future observations are  $(\tilde x, \tilde y)$.    In a predictive paradigm \citep {bernardo2001bayesian, Vehtari+Ojanen:2012},  the statistical inference should be inference on observable quantities such as the future observation $\tilde y$, where  Bayesian decision theory gives a natural framework for the prediction evaluation. Therefore, we can view model comparison, model selection, as well as model combination as formal Bayesian decision problems.   At a higher level, whether to make a single model selection or model combination is part of the decision. 
 
 Given a model $M$ with its  parameter vector $\theta$, we  compute the posterior predictive density  $p(\tilde y|y, M)= \int \!p(\tilde y|\theta, M)p(\theta |y, M) d\theta$, where we have suppressed the dependence on $x$  for brevity.  To evaluate how close the prediction is to the \emph{truth},  we construct  the utility function of the  predictive  performance through \emph{scoring rules}.    In general, conditioning on $\tilde x$,  the unobserved future outcome  $\tilde y$ is the random variable in sample space $( \mathcal{Y}, \mathcal{A})$. $\mathcal{P}$ is a convex class of probability measure on $\mathcal{Y}$.  Any member of $\mathcal{P}$ is called a probabilistic forecast.   A scoring rule \citep{score} is a function $S:  \mathcal{P} \times  \mathcal{Y} \to [\infty, \infty]  $  such that $S(P, \cdot)$ is  $\mathcal{P}$-quasi-integrable for all  $P \in \mathcal{P}$.  In the continuous case,  every distribution $P\in \mathcal{P}$ is identified with its density function $p$.
 
 For two probability measures $P$ and $Q$,  we write $S(P, Q)=\int_{\mathcal{Y}}S(P, \omega) dQ(\omega) $.  A scoring rule $S$ is called \emph{proper} if $S(Q,Q) \geq S(P,Q)$  and \emph{strictly proper} if equality holds only when $P= Q $ almost surely.   A proper scoring rule defines the divergence $d: \mathcal{P} \times \mathcal{P} \to [0, \infty) $ as  $d(P,Q)=S(Q,Q)-S(P,Q)$.  
 For continuous variables, some popularly  used scoring rules include:
 \begin{itemize}
 	\item {\em Quadratic score:}  $\mathrm{QS}(p,\tilde y) = 2p(\tilde y)-||p||_2^2$ with the divergence $d(p,q)= ||p-q ||_2^2$.
 	
 	\item {\em Logarithmic score:}
 	$\mathrm{LogS}(p,\tilde y)=\log p(\tilde y) $ with $d(p,q)=\mathrm{KL}(q,p).$
 	The logarithmic score is the only proper local score assuming regularity conditions. 
 	
 	\item {\em Continuous-ranked probability score:}
 	$\mathrm{CRPS}(F,\tilde y)=-\int_{\rm I\!R} (F(\tilde y')  - 1(y' \geq \tilde y) ) ^2 dy'$ with  $d(F, G)=\int_{\rm I\!R} (F(\tilde y)-G(\tilde y))^2 d\tilde y $, where $F$ and $G$ are the corresponding distribution functions.
 	
 	\item {\em Energy score:}
 	$\mathrm{ES}(P,y)= \frac{1}{2}\mathbb{E}_P || Y-Y' ||_2^\beta-  \mathbb{E}_P|| Y-y||_2^\beta $, where $Y$ and $Y'$ are two independent random variables  from distribution $P$.
 	When $\beta=2$, this becomes $\mathrm{ES}(P,\tilde y)= -||\mathbb{E}_P(\tilde y)-\tilde y||^2.$
 	The energy score is strictly proper when $\beta \in (0,2)$ but not when $\beta=2$.  
 	
 	\item {\em  Scoring rules depending on first and second moments:}   Examples include $S(P, \tilde y)=-\log\mathrm{det} (\Sigma_P)- (\tilde y- \mu_P )^T   \Sigma_p^{-1}(y-\mu_P )$, where $\mu_P $ and $\Sigma_P$ are the mean vector and covariance matrix of distribution $P$.
 \end{itemize}

 In such framework,  the  expected utility for any posterior predictive distribution $p(\cdot)$ is 
 \begin{equation}\label{eq_integral}
\mathbb{E}_{\tilde y} S\left( p(\cdot), \tilde y\right)  =  \int_{\mathcal{Y}}S(p, \tilde y)  p_t (\tilde y|y) d\tilde y ,
  \end{equation}
 where $p_t (\tilde y|y)$ is the unknown true data generating density of  outcomes $\tilde y$ given current observations. 

With  the widely used  logarithm score,  the expected log predictive density (elpd) of model $M$ is
  \begin{equation}\label{eq_elpd}
\text{elpd} =  \int_{\mathcal{Y}} \log p(\tilde y| y, M)   p_t (\tilde y|y) d \tilde y.
 \end{equation}
 
The general decision problem is an optimization problem that maximizes the expected utility within some decision space  $\mathcal{P}$: $p^{\mathrm{opt}}=\arg\max_{p\in \mathcal{P}}  \int S(p, \tilde y) d  p_t (\tilde y).$  \emph{Model selection} can   be viewed as a sub decision space of \emph{model combination}, by restricting model weights  to have only one non-zero entry.   In such sense, model selection may be  unstable and wasteful of information.
 
The expected scoring rule \eqref{eq_integral}  depends on the  generating process of $\tilde y$, which is unknown in the first place.    How we will estimate such expectation depends on how we view the relation between  belief models and  the true generating process, i.e., three $\mathcal{M}$-views.

 \subsection{Remodeling: $\mathcal{M}$-closed,  $\mathcal{M}$-complete, and  $\mathcal{M}$-open views}\label{sec_M_views}  
\citet{bernardo2001bayesian}  classified  model comparison problems into three categories: $ \mathcal{M}$-closed, $ \mathcal{M}$-complete  and  $ \mathcal{M}$-open.

 \begin{itemize}
 	\item In $ \mathcal{M}$-closed problems,  the true data generating process can be expressed by one of $M_k \in \mathcal{M}$, although it is unknown to researchers.  
 	\item $\mathcal{M}$-complete refers to the situation where the true model exists and is out of model list $\mathcal{M}$. But we still wish to use a model $M^*$ because of  tractability of computations or communication of results, compared with the actual belief model.  
 	\item The  $\mathcal{M}$-open perspective acknowledges the  true model is not in $ \mathcal{M}$, and we cannot specify the explicit form $p(\tilde y|y)$ because it is too difficult conceptually or computationally, we lack time to do so, or do not have the expertise,  etc.   
 \end{itemize}

Computing the integral \eqref{eq_integral} requires a model for $\tilde y$.    The inference and model assessment can have different model assumptions,   akin to the distinction between estimation and hypothesis testing in frequentist statistics.
For $\mathcal{M}$-closed and $\mathcal{M}$-complete problems, we  specify a belief model $M^{*}$ that  we believe to be or well approximate the data generate process, and we describe all  uncertainty related to future data in the belief model $M^{*}$ through $p(\tilde y|  y, M^{*})$.  The expected utility of any  prediction $Q$ is estimated by  
\begin{equation}\label{eq_est_M_closed}
\mathbb{E}_{\tilde y} S(Q, \tilde y) \approx  \int_{ \mathcal{Y}} S(Q, \tilde y) p(\tilde y|  y, M^{*})d \tilde y.
\end{equation}
 $\mathcal{M}$-closed and $\mathcal{M}$-complete  are a simplification of reality.
No matter how flexible the belief model $M^{*}$ is,  there is little reason to believe it reflects the \emph{truth}, unless in rare situations such as computer simulations.  Although such simplification is sometimes useful,   the stronger assumption may also result in an unverifiable and irretrievably  bias in \eqref{eq_integral}, which will further lead to an undesired performance in model aggregation.

In $\mathcal{M}$-open problems,  we still  rely on models in $\mathcal{M}$ in inference and prediction.   But we make minimal  assumptions  in the model  assessment phase.  Cross-validation is a widely used strategy to this end, where we  re-use samples $y_1, \dots, y_n$ as pseudo Monte Carlo draws from the true data generating process without having to model it explicitly.  For example, the leave-one-out (LOO) predictive density of a model $M$ is a consistent estimation of \eqref{eq_elpd}.
$$
\text{elpd}_{\text{loo}} = \frac{1}{n} \sum_{i=1}^n \log p(y_i | y_{-i},  M)=   \frac{1}{n}\sum_{i=1}^n  \int\! \log p(y_i | \theta,  M)p(\theta| M, y_{1}, \dots, y_{i-1}, y_{i+1}, \dots, y_n )d\theta.
$$

\section{From Bayesian Model Averaging To Bayesian Stacking}\label{sec_compare}
We have a series of models $\mathcal{M}= \{M_1, \dots, M_K\}$, each having  parameter vectors $\theta_k\in \Theta_k$. In general $\theta_k$ can have different dimensions and interpretations, and some may be infinite dimensional too.  We denote the likelihood and prior in the $k$-th model by  $p(y|\theta_k)$ and  $p(\theta_k|M_k)$. The goal is to aggregate all  component predictive distributions $\{p(\tilde y | y, M), M\in\mathcal{M} \}$. Adopting different $\mathcal{M}$-views, we will solve the problem by various methods as follows. 

\subsection{$\mathcal{M}$-Closed: Bayesian Model Averaging}\label{sec_BMA}
Bayesian model averaging (BMA) assigns  a prior both to the model space  $p(M_k)$ and parameters $p(\theta_k|M_k)$. Through Bayes rule, the posterior probability of model $k$ is proportional to the product of its prior and marginal likelihood,
$$p(M_k | y)  =\frac{p(y | M_k ) p(M_k)} {\sum_{k^\prime=1}^K p(y | M_k^\prime ) p(M_k^\prime)} .$$
In particular, the aggregated posterior predictive distribution of new data $\tilde y$ is estimated by
$$p_{\mathrm{BMA}}(\tilde y  | y)= \sum_{k=1}^K p(\tilde y| M_k, y)p(M_k |y).$$

 
In   $\mathcal{M}$-closed cases,  BMA is  optimal  if the method is evaluated  based on its frequency properties assessed over the joint prior distribution of the models and their internal parameters \citep{madigan1996bayesian, hoeting1999}.  In $\mathcal{M}$-open and $\mathcal{M}$-complete cases,  BMA almost always asymptotically \emph{select}  the one single model on the list that is closest in Kullback-Leibler (KL) divergence, compromising the  extra expressiveness of model aggregation.

Furthermore, BMA is contingent on the marginal likelihood $p(y | M_k ) = \int p(y | \theta_k) p(\theta_k| M_k)d \theta_k$,  which will be sensitive to the  prior $p(\theta_k | M_k)$.  A correct specification of the model (an $\mathcal{M}$-closed view) is stronger than the  asymptotic convergence to truth in some model, as it also requires  the prior to be correctly chosen in terms of reflecting the  actual population distribution of the underlying parameter.   For example, consider observations $y_{1}, \dots, y_{n}$ generated from $y\sim \N(0, 0.1^2)$, and a  normal-normal model: $y\sim \N(\mu, 1)$ with a prior $\mu\sim \N(0,10^2)$. Such prior is effectively flat on the range of observed $y$.  However,  a  change of prior to $\mu\sim \N(0,100^2)$ or $\N(0,1000^2)$ would divide the marginal likelihood, and thereby the posterior probability,  by roughly a factor of $10$ or $100$.

\subsection{$\mathcal{M}$-Open:  Stacking}\label{sec_stacking}

Stacking is originated from machine learning for the purpose of pooling point estimates from multiple regression models \citep{wolpert1992,breiman1996,leblanc1996}. \citet{clyde2013bayesian}, \citet{le2016bayes}, and \citet{yao2018using} develop and extend its   Bayesian interpretation.

The ultimate goal of stacking a set of $K$ predictive distributions built from the model  list $\mathcal{M}= ( M_1, \dots, M_K )$ is to find the  predictive distribution with the form of a linear pooling  $\mathcal{C}= \{\sum_{k=1}^{K}w_k p(\cdot | M_k):   \sum_k w_k=1, w_k \geq 0  \}$ that is optimal according to a specified utility.  The decision to make is the model weights $w$, which has to be a length-$K$ simplex $w\in \mathbb{S}_1^K  = \{ w\in[0,1]^K: \sum_{k=1}^K w_k = 1\}$.      Given a scoring rule $S$, or equivalently the divergence  $d$, the optimal stacking weight should solve
\begin{equation} \label{stacking_population}  \max_{w \in \mathbb{S}_1^K}  S\Bigl(   \sum_{k=1}^K w_k p(\cdot | y, M_k), p_t(\cdot | y) \Bigr)\;\;       \text{or equvalently}\;  \min_{w \in \mathbb{S}_1^K}  d \Bigl( \sum_{k=1}^K w_k p( \cdot | y, M_k ), p_t(\cdot | y) \Bigr),   
\end{equation}
where $ p(\tilde{y} | y, M_k)$ is the predictive density of new data $\tilde y$ in model $M_k$ that has been trained on observed data $y$ and $ p_t(\tilde{y}| y )$ refers to the true distribution.

With an $\mathcal{M}$-open view, we   empirically estimate the optimal stacking weight in  \eqref{stacking_population} by replacing the full predictive distribution  $ p(\tilde{y} | y, M_k)$  evaluated at a new data point $\tilde{y}$ with the corresponding LOO predictive distribution $\hat p_{k, -i}(y_i)= \int\! p(y_i | \theta_k, M_k) p(\theta_k | y_{-i}, M_k)  d\theta_k$. 

Therefore, it suffices to solve the following optimization problem 
\begin{equation} \label{stacking}
\hat w^{\mathrm{stacking}}=\max_{w \in \mathbb{S}_1^K }\frac{1}{n}\sum_{i=1}^n S\Bigl( \sum_{k=1}^K  w_k \hat p_{k,-i}, y_i\Bigr). 
\end{equation}
The aggregated predictive distributions on new data $\tilde y$ is   $ p_{\text{stacking}}(\tilde y |y)= \sum_{k=1}^K \hat w_k^{\mathrm{stacking}}  p(\tilde y|y, M_k)$.

In terms of \citet[][Section 3.3]{Vehtari+Ojanen:2012}, stacking  predictive distributions  \eqref{stacking} is the $M^{*}$-optimal projection of the information in the actual belief model $M^{*}$ to $\hat{w}$, where explicit specification of $M^{*}$ is avoided by re-using data as a proxy for the predictive distribution of the actual belief model and the weights $w_k$ are the free parameters.

\paragraph{Choice of utility.}
The choice of the scoring rule should depend on the underlying application and researchers'  interest. Generally we recommend  
  logarithmic score because (a)  log score is the only proper local scoring rule,  and (b) the easy interpretation of the underlying Kullback-Leibler divergence.    When using logarithmic score we name \eqref{stacking} as \emph{stacking of predictive distributions}:
\begin{equation}\label{stacking_log}
\max_{w \in \mathbb{S}_1^K } \frac{1}{n} \sum_{i=1}^n \log \sum_{k=1}^K w_k p(y_i | y_{-i}, M_k).
\end{equation}

\subsection{$\mathcal{M}$-Complete:  Reference-model Stacking} 
It is  possible to replace cross-validation with  a nonparametric reference model $M^*$.  Plug it into $\eqref{eq_est_M_closed}$ we  compute the expected utility  and further optimize over stacking weights, which we will call reference-model stacking.  
We can either stack component models $p(\tilde y|M_k)$, or stack the projected component models using a projection predictive approach which projects the information from the reference model to the restricted models \citep {piironen2017comparison}.
However in  general it is challenging to construct a useful reference model,   as then  there is probably no need for model averaging.

\subsection{The Connection Between BMA and Stacking}\label{sec_connect}
BMA, and more generally marginal likelihood based model evaluation,  can also be viewed as a special case of the utility-based model assessment. 

First, under an  $\mathcal{M}$-closed view, we believe the data is generated from one of the model $M^*\in \mathcal{M}$ in the candidate model list. We consider a zero-one utility by an indicator function of whether the model has been specified correctly:
\begin{equation}\label{eq_zero}
u(M^*, M_k)=\mathbbm{1}(M^* =  M_k). 
\end{equation}
Then the expected utility  $M_k$  is
$\int \mathbbm{1}(M^* =  M_k) p(M^* |y)d M^* = p(M_k|y)$, which is exactly  the posterior model probability  $p(M_k|y)$ in BMA.    Hence the decision maker will pick the model with the largest posterior probability, which is equivalent to the approach of Bayes factor.   Interestingly, the model with the largest BMA weight  is also the model to be selected under the zero-one utility, whereas in general the model with the largest stacking weight is not necessarily single-model-selection optimal (see discussions in Section \ref{sec_conclude})

Second,  under the $\mathcal{M}$-closed view 
the  information about  unknownness  is contained in the posterior distribution $p(M_k,\theta_k|y)$, and  the actual beliefs about the future observations are described by the BMA predictive distribution.  Using \eqref{eq_est_M_closed} and \eqref{stacking_population},  stacking over the logarithmic score reads
$$\max_{w\in \mathbb {S}_1^K} \int_{\mathcal{Y}} \log (    \sum_{k\prime=1}^K w_{k\prime} p(\tilde y|  M_{k\prime},  y )      )  \sum_{k=1}^K  p(M_k| y)  p(\tilde y|  M_k, y) d\tilde y, $$
whose optimal solution is always the same as the BMA weight $w^{\mathrm{opt}}_{k}=p(M_k| y)$,  as the logarithmic score  is strictly proper. 

In practice it is nearly impossible to either come up with an exhaustive list of possible candidate models that encompasses the true data generating process, or to formulate  the true prior that reflects the population.     It is not surprising that  stacking typically outperforms  BMA  in various prediction tasks \citep[see extensive simulations in][]{clarke2003, yao2018using}.    Notably, in the large sample limit, BMA  assigns weight $1$ to  the closest model to the true data generating process measured in KL divergence, regardless of how close other slightly more wrong models are. It effectively becomes model selection and  yields practically spurious and overconfident results \citep[e.g.,][]{yang2018good}  in $\mathcal{M}$-open problems.

\subsection{Hierarchical Stacking} \label{sec_hierarchical}
 
Model averaging is more likely to be useful when candidate models are more dissimilar---different models perform better or worse in different subsets of data. 
This suggests we can further improve the aggregated prediction by identifying which model can apply to which part of data,  
so that model averaging is a step toward model improvement rather than an end to itself.  

Hierarchical stacking \citep{yao2021bayesian} allows the model weight $w$ to vary by input covariate $x$, such that at any  input location $\tilde x\in \mathcal{X}$, the ``local'' model weight $w(\tilde x)$ is a length-$K$ simplex vector.   The aggregated conditional prediction becomes $p(\tilde y| \tilde x, w) = \sum_{k=1}^K     \hat w_k(\tilde x) p(\tilde y | \tilde x, M_k)$.

For example, if $x$ is discrete and takes on $J$ different values in the data, we need to construct a $J\! \times\! K$ matrix of weights  such that $w(x=j)= w_{jk}$, which can be mapped to an unconstrained weight space  $\alpha_{jk} \in \R^{J(K-1) }$ via softmax:
$$
	w_{jk}= \frac{\exp(\alpha_{jk})}{  \sum_{k=1}^K \exp( {\alpha_{jk} } )},  ~ 1\leq k\leq K-1,~ 1\leq j \leq J;  \qquad \alpha_{jK}=0,   ~1\leq j\leq J.
$$
Because of the larger decision space,  separately solving stacking  \eqref{stacking}  for all $j$  leads to large variance.  To partially pool the local weights  across $x$,  we can  use a hierarchical prior  conditional on hyperparameters $\mu\in\R^{K-1}$ and $\sigma\in\R_+^{K-1}$,
$$
\begin{gathered}
	\mathrm{prior:} \quad \alpha_{jk} \mid \mu_k, \sigma_k \sim \mbox{normal} (\mu_k, \sigma_k),  ~ k=1, \dots, K-1, ~ j=1, \dots, J,\\
	\mathrm{hyperprior:} \quad    \mu_k\sim  \mbox{normal}(\mu_0, \tau_\mu),  \quad \sigma_k \sim  \mbox{normal}^+(0, \tau_\sigma), \quad  k=1, \dots, K-1.
\end{gathered}
$$
Hierarchical stacking then folds the model averaging task into a hierarchical Bayesian inference problem.  Up to a normalization constant,  the  log joint posterior density of all free parameters $\alpha\in \R^{J\times K}, \mu\in \R^{K-1}, \sigma\in \R_{+}^{K-1}$ is defined by
$$
	\log p(\alpha, \mu, \sigma| \mathcal{D}) = \sum_{i=1}^n \log  \left( \sum_{k=1}^K w_k(x_i) \hat p_{k,-i} (y_i)\right) + \sum_{k=1}^{K-1}\sum_{j=1}^J \log p^{\mathrm{prior}} \left( \alpha_{jk}|  \mu_k, \sigma_k \right) +  \sum_{k=1}^{K-1}\log p^ {\text{hyper}\atop\text{prior}} \left( \mu_k, \sigma_k \right).
$$
This formulation generalizes  log-score-stacking \eqref{stacking},  as the latter method equals  the maximum-a-posteriori (MAP) solution of hierarchical stacking when all $\sigma_k =  0$. \citet{yao2021bayesian}  discuss other extensions of hierarchical stacking,  including regression for continuous predictors, nonexchangeable models for nested or crossed grouping factors, and nonparametric priors.

\subsection{Other Related Methods and Generalizations}
The methods above have multiple variants.  

When the marginal likelihood in BMA is hard to evaluate, it can be approximated by information criterion.  In  Pseudo Bayes factors \citep{ Geisser+Eddy:1979, Gelfand:1996}, we replace the  marginal likelihoods $p(y|M_k)$ by a product of Bayesian leave-one-out cross-validation predictive densities $\prod_{i=1}^np(y_i | y_{-i}, M_k)$.  
 \citet{yao2018using} propose another  information criterion based weighting scheme named Pseudo-BMA weighting. The weight for model $k$ is  proportional to the exponential of the model's estimated elpd:  $w_k  \propto \exp( \mathrm{\widehat {elpd}}^k_{\mathrm{loo}} )$.  Alternatively, such quantity can be estimated using a nonparametric reference model in $\mathcal{M}$-complete views \citep{li2019comparing}.   We may further take into account the sampling variance in cross-validation, and average over weights in multiple Bayesian bootstrap resamples \citep{yao2018using}.   The information criterion weighting is computationally easier, but should only be viewed as an approximation to the more desired stacking weights.   

We may combine the cross-validation and BMA.  Intrinsic Bayesian model averaging \citep[iBMA, ][]{berger1996intrinsic} enables improper prior, which is not allowed in BMA. It first partitions samples into a small training set $y(l)$ and remaining $y(-l)$, and replaces the marginal likelihood by partial likelihood $\int\! p(y(-l)|M_k, \theta_k) p(\theta_k|y (l), M_k)d \theta$.  The final weight   is the average across some or all possible training samples.  An alternative is to avoid averaging over all subsets and use the fractional Bayes factor \citep{o1995fractional}. 
iBMA is more robust for models with vague priors, but is reported to underperform stacking.

All model aggregation techniques introduced so far are two-step procedures, where we first fit individual models and then combine all predictive distributions. It is also possible to conduct both steps jointly, which can be viewed as a decision problem on both the model weights and component predictive distributions.   Ideally, we may avoid the  model combination problem by extending the model to include the separate models $M_k$ as special cases.   A finite-component mixture model is the easiest model expansion, but is generally quite expensive to make inference. Further, if the sample size is small or several components in the mixture could do the same thing, the mixture model can face non-identification or instability.  In fact, the immunity to duplicate models is a unique feature of stacking, while many methods including BMA,  information criterion weighting and mixture models often have a  disastrous performance in face of  many similar weak models.

Apart from combining  \emph{models},  when we  fit one single model but unstable computation,  model averaging techniques are also useful to combine \emph{inference} results from  multiple non-mixing runs. This is related to the idea of bagging \citep{breiman1996bagging}.
In particular, when the posterior density  $p(\theta|y)$ from a  model contains multiple isolated modes,   Markov chain Monte Carlo (MCMC) algorithms can  have difficulty moving between modes.  \citet{yao2020stacking} propose to use parallel runs of randomly-initialized MCMC, variational, or mode-based inference to hit as many modes or separated regions as possible, and then reweigh and combine the  posterior Monte Carlo draws using stacking \eqref{stacking}. 
The result from multi-run stacking is not necessarily equivalent, even asymptotically, to fully Bayesian inference, but it serves many of the same goals. With a misspecified model and multimodal posterior density, multi-run stacking could lead to better predictive performance than the full Bayesian inference.

\section{Asymptotic Theories of Stacking}\label{sec_theory}
To better understand how stacking works, we outline three theory properties in the following.
\subsection{Model Aggregation Is No Worse Than Model Selection}\label{sec_consistent}
The stacking estimate  (\ref{stacking_population}) finds the optimal predictive distribution within the linear combination that is the closest  to the data generating process with respect to the chosen scoring rule.   Solving for the stacking weights in (\ref{stacking_log})  is an M-estimation problem.   To what extent shall we worry about the finite sample error in leave-one-out cross-validation?
 Roughly speaking,  as long as there is   consistency for single model cross-validation,   then  asymptotically model averaging never does  worse than model selection in terms of prediction \citep{clarke2001combining}.  \citet{le2016bayes} further prove that under some mild conditions,  for either the logarithmic scoring rule or the energy score (negative squared error)  and a given set of weights $w_1 \dots w_K$, the weighted leave-one-out-score is a consistent estimate as sample size $n \to \infty$,   
$$
\frac{1}{n}\sum_{i=1}^n S\Bigl( \sum_{k=1}^K  w_k \hat p_{k,  -i}  , y_i \Bigr) -  \mathbb{E}_{\tilde y| y} S\Bigl(  \sum_{k=1}^K w_k p(\tilde y| y , M_k) , \tilde y\Bigr)          \xlongrightarrow{\text{$L_2$}}   0.
$$
In this sense,  stacking gives optimal combination weights asymptotically, and is an approximation to the Bayes action.

\subsection{Stacking  Viewed as Pointwise Model Selection}\label{sec_novel}
Besides justified by the decision theory, stacking weights also have a probabilistic interpretation.  To see this,  we  divide the input-output product space $\mathcal{X}\times \mathcal{Y}$ into $K$ disjoints subsets based on which model performs the best, 
$$\mathcal{J}_k \coloneqq  \{(\tilde x, \tilde y)\in \mathcal{X}\times \mathcal{Y}: p(\tilde y|M_k, \tilde x)> p(\tilde y|M_{k\prime}, \tilde x),  \forall k^\prime\neq k\}. ~ k=1, \dots, K.$$ 
We call a family of predictive densities $\{p(\tilde y| M_k, \tilde x)\}_{k=1}^K$ to be locally separable with a constant pair  $L>0$ and  $0\leq \epsilon< 1$, with respect to the true data generating process $p_t(\tilde y,  \tilde x)$, if
\begin{equation}\label{eq_sep_y}
	\sum_{k=1}^K \int_{(\tilde x, \tilde y)\in \mathcal{J}_k  } \mathbbm{1}\Big(  \log p(\tilde y|M_k, \tilde x)<   \log p(\tilde y|M_{k^{\prime}}, \tilde x)+ L,  \; \forall    k^{\prime} \neq k \Big)  p_t(\tilde y,  \tilde x) d\tilde y d\tilde x \leq \epsilon.
\end{equation}

\citet{yao2021bayesian} show that under the separation condition \eqref{eq_sep_y}, the log score stacking weight \eqref{stacking}  is  approximately  the probability of the model being the locally best fit:  $
		w^{\mathrm{stacking}}_{ k}  \approx   \Pr(\mathcal{J}_k)$, where the probability is taken  with respect to the joint true data generating process.


\subsection{Selection or Averaging?}\label{sec_conclude}
The advantage of model averaging comes from the fact that model can behave differently in different regions in $(x, y)$ space.    Let $\rho=\sup_{k} \Pr(\mathcal{J}_k)$, then  $1- \rho$ is a rough description of the  diversity of models.  In terms of the  expected log predictive density (elpd),  \citet{yao2021bayesian} show that under the  separation condition \eqref{eq_sep_y}, the gain from the optimally weighted 
models (against model selection) is lower bounded by  
$$\mathrm{elpd}_{\mathrm{stacking}} - \sup_k \mathrm {elpd}_k \geq    L(1-\rho)(1-\epsilon)- \log K.$$

One practical difficulty in model comparison is to determine how large the difference between  model performance is ``significant"  and whether to discard bad models \citep{sivula2020uncertainty}. The probabilistic approximation in the previous subsection suggests that an overall weak model can still be useful in the  aggregation.  As long as a model is  better than all remaining models in some subset of data, this model  possesses a non-zero stacking weight  no matter how  poorly it fits everywhere else.

Lastly,  a model with the largest BMA weight (assuming equal prior) is  optimal under marginal likelihood model selection.   In contrast,  a model with the largest stacking weight is not necessarily optimal in terms of single model selection:  it may outperform other models most of the time but also have arbitrarily low elpd in the remaining areas---stacking is not designed for model selection. Hence,  we do not recommend to discarded models with small  weights from the average.

\section{Stacking in  Practice}\label{sec_practice}

\subsection{Practical Implementation Using Pareto Smoothed Importance Sampling}
Stacking (\ref{stacking}) requires leave-one-out (LOO) predictive density $p(y_i | y_{-i}, M_k)$,  whose exact evaluation needs to refit  each model $n$ times.    $k$-fold cross-validation is computationally cheaper but may introduce higher bias. \citet{vehtari2017practical} proposed the an approximate method for Bayesian LOO.  It is based on the importance sampling identity:
$$p(\theta | y_{-i}) \propto \frac{1}{p(\theta | y_i)} p(\theta  |  y_1, \dots, y_n).$$
In the $k$-th model, we fit to all the data, obtaining $S$ simulation draws $\theta_k^s (s=1,\dots S)$ from the full posterior $p(\theta_k|y, M_k)$  and  calculate 
\begin{equation} \label{ratio}
r_{i,k}^s =\frac {1} {p(y_i | \theta^s_k, M_k) } \propto \frac{p(\theta^s_k | y_{-i}, M_k)}{ p(\theta^s_k | y,  M_k) }. 
\end{equation}
A direct importance sampling often has high or infinite variance and we remedy it by  Pareto smoothed importance sampling \citep[PSIS, ][]{vehtari2015pareto}.  For each fixed model $k$ and data $y_i$, we fit the generalized Pareto distribution to a set of largest importance ratios $r_{i,k}^s$,  and calculate the expected values of the order statistics of the fitted generalized Pareto distribution. These values are used to obtain the smoothed importance weight $w_{i,k}^s$, which is used to replace $ r_{i,k}^s$.  PSIS-LOO importance sampling  computes the LOO predictive density as 
\begin{align*}\label{elppd}
p(y_i | y_{-i}, M_k )  =  \int\!p(y_i |\theta_k, M_k)  \frac{p(\theta_k | y_{-i}, M_k)}{p(\theta_k | y, M_k) }  p(\theta_k| y, M_k) d \theta_k 
\approx  \frac{ \sum _{s=1}^S w_{i,k}^s p(y_i | \theta_k^s, M_k)   }{  \sum _{s=1}^S w_{i,k}^s }.
\end{align*}

An \texttt{R} package \texttt{loo} \citep{loopackage} provides model weights from the PSIS-LOO  based stacking  and pseudo-BMA. Suppose \texttt{fit1}, \texttt{fit2} and \texttt{fit3} are three  models fit  objects from the Bayesian inference package Stan \citep{Stan}, then we can compute their stacking weights as follows.
\begin{verbatim}
model_list <- list(fit1, fit2, fit3)
log_lik_list <- lapply(model_list, extract_log_lik)
# stacking:
wts <- loo_model_weights( log_lik_list, method = "stacking", 
optim_control = list(reltol=1e-10))
\end{verbatim}


\subsection{Stacking for Multilevel Data}
Although the illustration in this article is focused on iid data,  the   leave-one-out  consistency only requires the conditional \emph{exchangeability}  of outcomes $y$ given $x$ \citep[][Chapter 6]{bernardo2001bayesian}.   \citet{roberts2017cross} review cross‐validation strategies for data with temporal, spatial, hierarchical, and phylogenetic structure. 
In general, the PSIS-LOO approximation applies to factorizable models $p(y|\theta, x)= \prod_{i=1}^N p(y_i|\theta, x_i)$ such that the pointwise log-likelihood can be obtained easily by computing $\log p(y_i|\theta, x_i)$.

Non-factorizable models can sometimes be factorized  by re-parametrization. In a multilevel model with $J$ groups,  we denote the group-level  and global parameter as $\theta_m$ and $\psi$.  The joint likelihood is
\begin{equation}\label{multi-level}
p(y|x, \theta, \psi)= \prod_{j=1}^J \left[ \prod_{n=1}^{N_j} p(y_{jn}|x_{jn},\theta_j) p(\theta_j|\psi)  \right] p(\psi),
\end{equation}
where $y$ are partially exchangeable, i.e., $y_{mn}$ are exchangeable in group $j$, and $\theta_m$ are exchangeable. 
Rearrange the data and denote the group label of $(x_i, y_i)$ by $z_i$, then \eqref{multi-level} can be reorganized into the long format $\prod_{i=1}^{N'} p(y_i|x_i, z_i,\theta, \psi)$ so the previous results follow.  Depending on whether the prediction task is to predict a new observation within a  group, or a new group, we should consider leave-one-point-out or leave-one-group-out cross-validation. 

When the future data are known to come from a  group $j$,  there are two stacking strategies:  (a) apply  generic stacking  only to  observations from the $j$-th group, which is asymptotically optimal with enough data, but has large variance if the group size is small, and (b) apply stacking to all observations  regardless of their group structure, which has  smaller variance at the cost of less flexibility.    The more preferred hierarchical stacking (Section  \ref{sec_hierarchical}) trades off between these two extremes. 
Its Bayesian hierarchical formulation shares information across groups, stabilizing model weights in small groups while still allowing the flexibility of group-specific weighting.

\subsection{Stacking for Time Series Data}
When observations $y_t$ come in sequence and  the main purpose is to make prediction for the next not-yet-observed data, we can use  the prequential principle \citep{dawid1984present} to factorize the likelihood:
$p(y_{1:N}| \theta) =  \prod_{t=1}^N p(y_t | y_{1:t-1} , \theta ).$
In model averaging,  we can replace the LOO density $p(y_i| y_{-i})$ in \eqref{stacking} by the sequential predictive density leaving out all future data: $p(y_t | y_{<t})=  \int\! p(y_t | y_{1:t-1}, \theta ) p (\theta| y_{1:t-1}) d\theta$ in each model, and then stacking follows.  
The ergodicity of $y$ will yield
$$
\lim_{N\to \infty} \frac{1}{N}\sum_{t=1}^N S \left(   p(\cdot | y_{<t} ) , y_t \right) -   \lim_{N\to \infty} \frac{1}{N}  \mathbb{E}_{Y_{1:N}}  \sum_{t=1}^N     S \left(   p(\cdot | Y_{<t} ) , Y_t \right)  \to 0.
$$
which implies a similar  stacking optimality  as discussed in Section  \ref{sec_consistent}.  \citet{geweke2012prediction} investigate this stacking approach in time series data.

When there is a particular horizon of interest for prediction, a model that is good at short term forecast is not necessarily good for long term forecast.  We can extend the one-step ahead   $p(y_t |y_{<t})$  to  $m$-step-ahead predictive density  $p(y_{t<m}| y_{<t}) = p(y_t,\ldots,y_{t+m-1}|y_{1},\ldots,y_{t-1})= \int\! p(y_{t<m}|y_{<t}, \theta)p(\theta|y_{<t})d\theta$ in the objective function \citep{lavine2019adaptive}.

In terms of computation, the exact prequential evaluation requires refitting  each model  for each $t$, which can be approximated  by PSIS as,
$p(y_t | y_{<t})  =  \int\!p(y_t |\theta, y_{<t}) \frac{p(\theta | y_{<t} )}{p(\theta|  y) }  p(\theta| y) d \theta. $
We then start from the full data inference $p(\theta| y)$ and dynamically update $p(\theta | y_{<t} )$ using PSIS approximation. When  $p(\theta | y_{<t} )$ reveals large discrepancy from  $p(\theta|  y)$ for some small $t$,  we refit the model $p(\theta | y_{<t} )$  and update the proposal. \citet{Buerkner2018SAP} verify such approximation gives stable and accurate results with minimal number of refits in time series.

We can further extend the static stacking scheme to  a dynamic model weighting, allowing the explanation power of models to change over time. \citet{yao2021bayesian} present an election  forecast example that applies  hierarchical stacking to longitudinal polling data. Another flexible  model weighting strategy in  time series forecasting is  Bayesian predictive synthesis \citep[BPS,][]{mcalinn2017dynamic, mcalinn2017multivariate}: The predictive density has the form $ \int\! \alpha(y|z) \prod_{k=1:K} h_{k}(z_k) d z$,  where $z = z_{1:K}$ is the latent vector generated from  predictive densities $h_k(\cdot)$ in each model and $\alpha(y|z)$ is the distribution for $y$ given $z$ that is designed to calibrate the model-specific biases and correlations.

 \subsection{The Choice of Model List} 
As we have discussed earlier, BMA and information criterion weighting are undesired against many similar weak models. We may remedy this by a careful construction of priors.  For example, \cite{george2010dilution} establishes dilution priors to compensate for model space redundancy in linear models, putting smaller weights on those models that are close to each other.  \cite{fokoue2011bias} introduce prequential model list selection to obtain an optimal model space. 

Stacking is prior invariant and immune to model duplication. Nevertheless, all methods discussed in the present paper fit models separately, and are thereby limited in that they do not pool information between the different model fits.  The benefit of stacking depends only on the span of the model list \citep{le2016bayes},  and models to be stacked should be as different as possible \citep{breiman1996}.  In light of discussion in Section \ref{sec_novel}, the ideal situation of stacking is when models can offer  different  predictive density pointwisely.

In general, we do not recommend constructing a large list of weak models (e.g., subset regression) and aggregate them in a black box way, as in that setting we would recommend moving to a continuous model space that encompasses all separate models. We prefer to  carefully construct  component models that would have individually fit the data as much as possible,   and all admissible estimators for parameters should be considered before the optimization procedures.
 

 \section{Discussion}
Along with an increasing number of  statistical models and learning algorithms, ensemble methods have been  appealing  tools to expand  existing models and inferential procedures,  and to improve predictive performance.   In addition, the popularity of ensemble methods  in Bayesian statistics can be viewed as representing a modern shift in Bayesian data analysis: from a static model-based inference to a Bayesian workflow in which we are fitting many models while working on a single problem.  

This article is mostly about Bayesian model averaging, stacking, and their variants. For these methods, the model weights are trained after model-specific inferences, and the cost of the former is  typically much smaller than the latter.  Another popular approach to construct  ensembles is to train each model and the model weight  simultaneously or iteratively, such as in boosting \citep{freund1997decision},  gradient boosting \citep{friedman2001greedy},   and mixture of  experts \citep{jacobs1991adaptive}. These methods are  computationally intensive for full-Bayesian inference, but  more useful to combine weak learners.   On the other hand, different ensemble methods can be further aggregated: for example,  to stack fits from BMA and  mixture of experts.

Many of these ensemble methods had limited usage until enough computational resources and efficient approximation became  available. Conversely,  many model averaging strategies also help solve difficulties in statistical computing. For example, bagging stabilizes otherwise non-robust  point estimates, and stacking can be used in multimodal posterior sampling. 
 
Looking forward, there are many open questions.  To name a few,  both BMA and stacking are restricted to a linear mixture form, would it be beneficial to consider other aggregation forms such as  convolution of predictions or a geometric bridge of predictive densities? Stacking often relies on some cross-validation, how can we better account for the finite sample variance therein?  While staking can be equipped with many other scoring rules,  what is the impact of  the scoring rule choice on the convergence rate and robustness?   Beyond current model aggregation tools,  can we develop an automated ensemble learner that could fully explore and expand the space of model classes---for example, using an autoregressive (AR) model and a moving-average (MA) model to learn an ARMA model?  We leave these directions  for  future investigation.

\bibliographystyle{apalike}
\small
\bibliography{Bayesian_aggregration}

\end{document}